\documentclass[reprint, amsmath,amssymb,aps,pre,onecolumn,notitlepage,superscriptaddress
]{revtex4-1}

\usepackage{graphicx}
\usepackage{siunitx}
\usepackage{courier}
\usepackage{setspace}
\usepackage{mathtools,amsmath,mathbbol}
\usepackage{amssymb}
\usepackage{float}
\usepackage{graphicx}
\usepackage{orcidlink}

\usepackage{bm}
\usepackage{hyperref}
\hypersetup{
  colorlinks   = true, 
  urlcolor     = blue, 
  linkcolor    = blue, 
  citecolor   = blue 
}

\begin{document}
\setstretch{1.2}

\preprint{APS/123-QED}
\title{Simulating the rheology of dense suspensions using pairwise formulation of contact, lubrication and Brownian forces}
\author{Xuan Li}
\affiliation{%
 School of Engineering, The University of Edinburgh, King's Buildings, Edinburgh EH9 3FG, United Kingdom}%
\author{John R. Royer}
\affiliation{%
 School of Physics and Astronomy, The University of Edinburgh, King's Buildings, Edinburgh EH9 3FD, United Kingdom}%
 \author{Christopher Ness}
 \email{chris.ness@ed.ac.uk}
\affiliation{%
 School of Engineering, The University of Edinburgh, King's Buildings, Edinburgh EH9 3FG, United Kingdom}%
\date{\today}

\begin{abstract}
Dense suspensions of solid particles in viscous liquid are ubiquitous in both industry and nature,
and there is a clear need for efficient numerical routines to simulate their rheology and microstructure.
Particles of micron size present a particular challenge:
at low shear rates colloidal interactions control their dynamics while at high rates granular-like contacts dominate.
While there are established particle-based simulation schemes for large-scale non-Brownian suspensions using only pairwise lubrication and contact forces,
common schemes for colloidal suspensions generally are more computationally costly and thus restricted to relatively small system sizes.
Here we present a minimal particle-based numerical model for dense colloidal suspensions which incorporates Brownian forces in pairwise form alongside contact and lubrication forces.
We show that this scheme reproduces key features of dense suspension rheology near the collodial-to-granular transition,
including both shear-thinning due to entropic forces at low rates and shear thickening at high rate due to contact formation.
This scheme is implemented in LAMMPS, a widely-used open source code for parallelized particle-based simulations,
with a runtime that scales linearly with the number of particles making it amenable for large-scale simulations.
\end{abstract}

\maketitle

\section{Introduction}
Dense suspensions of Brownian and non-Brownian solid particles in viscous liquid present intriguing flow properties, and understanding their rheology is a subject of both fundamental and technological relevance~\cite{ness2022physics,stickel2005fluid}.
Of particular interest are suspensions comprising particles with radius $a\approx$~\SI{1}{\micro\metre},
or more broadly in the range $0.1-$\SI{10}{\micro\metre}.
These are present in numerous applications, in all areas of food science~\cite{jambrak2010ultrasound} and consumer products, as well as across manufacturing and construction sectors~\cite{roussel2010steady} and indeed in many geophysical contexts~\cite{kostynick2022rheology}.
Often their physics are challenging because their Brownian diffusion time may be comparable to the processing or macroscopic timescales involved in their use,
so that they sit at the boundary of colloidal and granular systems~\cite{guy2015towards}.

Particle-based simulation offers a promising route to better understand the physics of these materials,
providing highly-resolved information complementary to what can be obtained by experiment.
With simultaneous access to particle trajectories and bulk rheology,
one might devise new micromechanical constitutive equations~\cite{gillissen2020constitutive}
or develop microstructural insight that could guide the future analysis of experimental data.
Numerical models might also be useful for exploring parameter space and systematically linking aspects of particle-level physics (friction~\cite{seto2013discontinuous}, adhesion~\cite{richards2020role}, roughness~\cite{lobry2019shear})
to bulk flow behaviour.
As a result, one might aim to optimise industrial processes such as mixing and extrusion or indeed to optimise the design of materials themselves through additives,
using insight gained through particle-based simulation.

Stokesian Dynamics (SD)~\cite{brady1988stokesian,banchio2003accelerated} is a computational method used to simulate the rheological behavior of colloidal and granular particles suspended in a viscous fluid,
addressing the special case of inertia-free flow where the Stokes number is zero~\cite{ermak1978brownian}.
The method involves balancing all of the forces on each particle by evaluating their velocities \emph{via} a grand mobility matrix containing information on the relative positions of every particle in the system.
Despite accurately capturing the long and short range hydrodynamic
interactions between particles, SD has not been adopted widely as
a predictive tool in applied and industrial settings in the same way as other particle-based simulation methods have,
due to the complexity of its implementation and its computational expense (notwithstanding recent developments that significantly speed it up~\cite{sierou2001accelerated,fiore2019fast}).

The discrete element method (DEM)~\cite{cundall1979discrete}, on the other hand, is a particle-based computational method (a variant of molecular dynamics) that is used widely to simulate the behavior of granular materials including powders, particles and grains taking into account their pairwise interactions.
In contrast to SD, DEM does not balance forces on each particle.
Instead inertia is present, and one simply sums the forces
and the resultant leads to an acceleration
that can be realised through a conventional time-stepping algorithm such as Velocity-Verlet.
This approach has proven to be useful for studying overdamped suspensions under shear flow nonetheless~\cite{trulsson2012transition, ness2015flow},
where one introduces short ranged lubrication forces and sets the Stokes number to be $\mathcal{O}(10^{-2})$ or smaller.
This approach is pragmatic in the sense that the physics associated with
flowing dense suspensions can be implemented in existing, widely-used codes with large user bases,
so that they have a clear path to adoption in engineering and other applied contexts.
To date there is not, to our knowledge, a discrete element method simulation that includes the relevant physics of dense suspensions at the colloidal-to-granular interface, accounting for short-ranged hydrodynamics, Brownian forces, and (frictional) particle-particle contacts.

Here we present a minimal particle-based simulation model for predicting the rheology
of dense Brownian and non-Brownian suspensions.
Our model comprises hydrodynamic lubrication, particle-particle contacts and Brownian forces. 
After first describing the model in detail,
we present some aspects of the effective interactions and diffusion that arise,
before giving a detailed account of the rheological predictions of the model.
The model reproduces well the main features of the experimentally observed rheology of dense suspensions,
namely a low shear rate plateau that gives way to shear thinning and 
later shear thickening as the shear rate is increased,
with the relative viscosity of the suspension increasing sharply with solids volume fraction and particle-particle friction coefficient.

\section{Methodology}
\begin{figure}
\includegraphics[trim = 0mm 0mm 0mm 0mm, clip,width=0.95\textwidth,page=1]{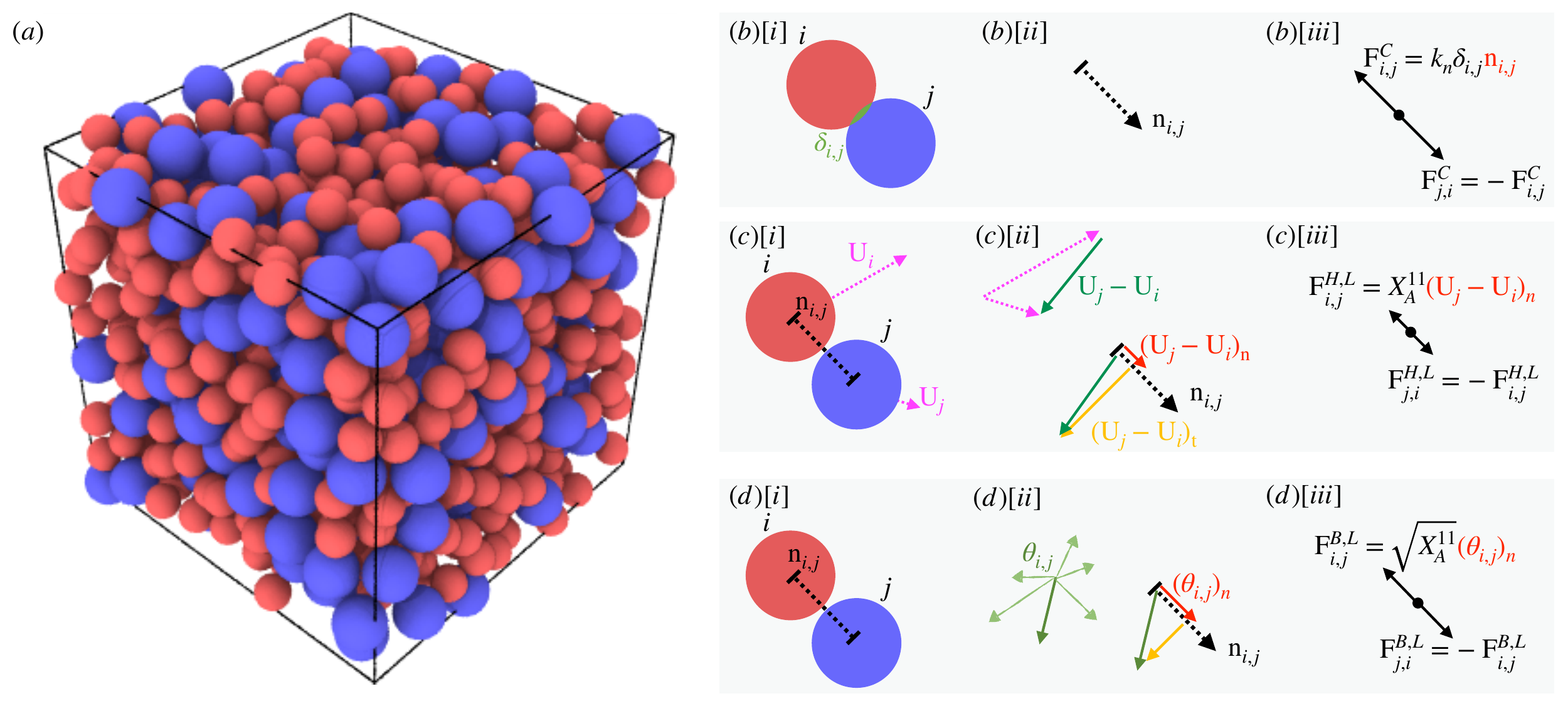}
\caption{
Snapshot of simulation box, and schematics of the leading pairwise force terms present in the model.
In all cases the leading component of the pairwise force
acts along the (positive or negative) direction of the unit vector $\bm{n}_{i,j}$
pointing from the centre of particle $i$ to the centre of particle $j$.
(a) Snapshot of the simulation box showing particles of radius $a$ (red) and $1.4a$ (blue). Particles are mixed in approximately equal volume.
(b) Contact force, with force acting along $\bm{n}_{i,j}$;
shown from left to right are
[i] sketch of contacting particles $i$ and $j$, with the overlap $\delta_{i,j}$ shown in green;
[ii] the unit vector $\bm{n}_{i,j}$ pointing from the centre of particle $i$ to $j$;
[iii] repulsive contact $F^C$ forces acting along the positive and negative directions of $\bm{n}_{i,j}$.
(c) Pairwise hydrodynamic lubrication force, with force set by the component of the relative particle velocity acting along $\bm{n}_{i,j}$;
Shown from left to right are
[i] sketch showing the velocity vectors $\bm{U}$ of neighbouring particles $i$ and $j$;
[ii] the relative velocity $U_j-U_i$ breaks down into tangential (yellow) and normal (red) components, with the latter pointing along $\bm{n}_{i,j}$;
[iii] lubrication forces act along the positive and negative directions of $\bm{n}_{i,j}$, proportional to the normal part of the relative velocity.
(d) Pairwise Brownian lubrication force with a random pairwise vector $\bm{\theta}_{i,j}$ projected onto $\bm{n}_{i,j}$.
Shown from left to right are
[i] neighbouring particles with centre-to-centre unit vector $\bm{n}_{i,j}$;
[ii] random vectors $\bm{\theta}_{i,j}$ drawn from a Gaussian distribution (green) are projected onto $\bm{n}_{i,j}$ by the tangential (yellow) and normal (red) operators;
[iii] Brownian forces act along the positive and negative directions of $\bm{n}_{i,j}$, proportional to the normal part of the random vector.
}
\label{figure1}
\end{figure}

We consider a model system of nearly-monodisperse solid spheres,
dispersed at high solids volume fraction $\phi$ in a density-matched Newtonian liquid.
The microscopic physics included in our model represent 
a minimal set of ingredients necessary to make
useful predictions of the rheology of suspensions comprising particles
with radius in the range $10^{-7}$ to $10^{-4}$m.
The trajectories of individual particles with translational and rotational motion are governed by Langevin equations that comprise three force ($\bm{F}$) and torque ($\bm{T}$) contributions:
direct particle contacts ($\bm{F}^\text{C}$, $\bm{T}^\text{C}$),
hydrodynamics ($\bm{F}^\text{H}$, $\bm{T}^\text{H}$),
and Brownian noise ($\bm{F}^\text{B}$, $\bm{T}^\text{B}$).
The equations of motion for translation and rotation of the particles are written, respectively, as
\begin{equation}
m_i\frac{d^2\bm{x}_i}{dt^2}
=
\sum_j\bm{F}_{i,j}^\text{C}
+
\bm{F}_{i}^\text{H,D}
+
\sum_j\bm{F}_{i,j}^\text{H,L}
+
\bm{F}_{i}^\text{B,D}
+
\sum_j\bm{F}_{i,j}^\text{B,L} \text{,}
\label{eqn:lang_force}
\end{equation}
\begin{equation}
\frac{2}{5}m_ia_i^2\frac{d\bm{\Omega}_i}{dt} = \sum_j\bm{T}_{i,j}^\text{C}
+
\bm{T}_{i,j}^\text{H,D}
+
\sum_j\bm{T}_{i,j}^\text{H,L}
+
\bm{T}_{i,j}^\text{B,D}
+
\sum_j\bm{T}_{i,j}^\text{B,L}
\text{,}
\label{eqn:lang_torque}
\end{equation}
where $\bm{x}_i$ represents the position of particle $i$,
$\bm{\Omega}_i$ represents its rotational velocity,
and $a_i$ and $m_i$ are its radius and mass respectively.
The subscript $i$ represents single-body forces and torques acting on particle $i$,
while the subscript $i,j$ represents pairwise forces and torques acting between particles labelled $i$ and $j$.
The superscripts C, H, B, D, L refer to
the force and torque components arising due to contacts (C), hydrodynamics (H)
and Brownian (B) effects, with the latter two acting both through drag (D) and lubrication (L).
Each of these force and torque terms is described in detail below.
These equations of motion can be understood as Langevin equations in which
the $\langle\cdot\rangle^C$ terms represent particle-particle interactions;
the $\langle\cdot\rangle^H$ terms represent configuration-dependent viscous friction (\emph{i.e.} dissipative forces linear in the particle velocities);
and the $\langle\cdot\rangle^B$ terms represent configuration-dependent (multiplicative) noise.
Although particle inertia is present in the model
we omit fluid inertia~\cite{hinch1975application},
arguing that for the regimes of interest the principle contributions to the overall bulk rheology
will come from particle-particle contact and hydrodynamic lubrication interactions.
Particles are subjected to a liquid flow field
given by $\bm{U}^\infty$ (acting through the body force $\bm{F}^{H,D}_i$ as described below),
leading to a rate of strain tensor $\mathbb{E}=\frac{1}{2}\left(\nabla\bm{U}^\infty + (\nabla\bm{U}^\infty)^\text{T}\right)$.
Pairwise forces and torques are summed over the neighbours $j$ of each particle $i$, and the positions,
velocities and acceleration are updated in a stepwise manner following the Velocity-Verlet algorithm~\footnote{We note that in LAMMPS the \emph{skin} argument of the \texttt{neighbour} command has units of [length].}. 
Below we describe each of the force and torque contributions in detail;
shown in Figure~\ref{figure1} are illustrative schematics of each of the forces.

\subsection{Contact forces and torques}
The particle-particle contact force $\bm{F}^C$ follows a conventional granular-type interaction~\cite{cundall1979discrete},
and is activated for any two particles $i$ and $j$ for which the centre-to-centre distance $|\bm{r}_{i,j}|$ is smaller than the sum of the radii $a_i + a_j$.
Contact forces include a repulsive part acting normal to the pairwise centre-to-centre vector $\bm{r}_{i,j}$
(we define a unit vector $\bm{n}_{i,j} = \bm{r}_{i,j}/|\bm{r}_{i,j}|$), and a tangential part.
For simplicity we model contacts as linear springs,
so that particle pairs experience repulsive contact forces proportional to their scalar overlap, defined once in contact as $\delta_{i,j} = (a_i+a_j)-|\bm{r}_{i,j}|$.
The implementation of our model within \texttt{LAMMPS}~\cite{plimpton1995fast}
nonetheless allows straightforward implementation of
more complex $\delta_{i,j}$ dependence.
Tangential forces are linear in $\bm{\xi}_{i,j}$,
a vector describing the accumulated displacement of the particle pair perpendicular to $\bm{n}_{i,j}$
since the initiation of the contact.
Contact force and torque magnitudes are controlled by normal and tangential stiffness constants $k_n$ and $k_t$ 
that set the hardness of the particles.
The force and torque are given respectively by:
\begin{equation}
    \bm{F}_{i,j}^\text{C}=k_n\delta_{i,j} \bm{n}_{i,j}-k_t\bm{\xi}_{i,j} \text{,}
    \label{eqn:force_contact}
\end{equation}
\begin{equation}
    \bm{T}_{i,j}^\text{C}=a_i(\bm{n}_{i,j} \times k_t\bm{\xi}_{i,j}) \text{.}
    \label{eqn:torque_contact}
\end{equation}
We additionally introduce a static friction coefficient $\mu$
that constrains the tangential force to $|k_t\bm{\xi}_{i,j}|\leq\mu k_n\delta_{i,j}$.
For larger values of $k_t\bm{\xi}_{i,j}$
the tangential part of the force and the torque are truncated,
and particle contacts transition from a rolling to a sliding regime.
We present data for $\mu=0$ throughout, except in Figure~\ref{figure6}(b)-(c) where we explore the role of contact friction.
Each pairwise contact between particles $i$ and $j$ contributes to the overall \emph{contact stress} of the system with a tensorial stresslet given by the outer product
$-\bm{F}^\text{C}_{i,j} \otimes \bm{r}_{i,j}$.
The contact stress $\mathbb{\Sigma}^C_{i,j}$ is obtained by summing this quantity over all contacting particle pairs and dividing by the system volume and dimension.

Contact forces of this kind have successfully been
deployed in numerical models for rate-independent granular suspension rheology~\cite{boyer2011unifying,cheal2018rheology,trulsson2012transition} and for models of shear thickening suspensions~\cite{seto2013discontinuous} (in the latter case rate dependence arises from a `critical load' that the contact force must exceed before static friction is activated).

\subsection{Hydrodynamic forces and torques}

In general,
hydrodynamic interactions in suspensions
appear as single particle drag forces $\bm{F}^\text{H,D}$,
pairwise near-contact lubrication forces $\bm{F}^\text{H,L}$,
and many-body long range forces.
In high volume fraction dense suspensions, however,
it is argued by many authors that the hydrodynamic interactions are dominated by near-contact lubrication interactions \cite{ball1997simulation} (which diverge on close approach) and that long range interactions are effectively \emph{screened} by intervening particles~\cite{seto2013discontinuous,more2020effect}. We follow this reasoning and therefore omit long range hydrodynamics from our model. Below we describe in detail the drag and lubrication forces deployed in the model. Single particle drag forces and torques are given by
\begin{equation}
    \bm{F}^\text{H,D}_{i}=6\pi\eta a_i (\bm{U}^\infty(\bm{x}_i)-\bm{U}_i) \text{,}
    \label{eqn:force_drag}
\end{equation}
\begin{equation}
    \bm{T}^\text{H,D}_{i}=8\pi\eta a^3_i (\bm{\Omega}^\infty-\bm{\Omega}_i) \text{,}
    \label{eqn:torque_drag}
\end{equation}
where we use the isolated-particle Stokes terms and,
for simplicity,
do not introduce volume fraction dependent hindrance functions.
Here $\eta$ is the liquid viscosity, $\bm{U}^\infty(\bm{x}_i)$ is the value of the liquid streaming velocity at the position of the centre of mass of particle $i$,
and $\bm{\Omega}^\infty = \frac{1}{2}\left(\nabla\times\bm{U}^\infty \right)$ (spatially uniform assing $\bm{U}^\infty$ is uniform in space).
The drag forces lead to a per particle stress given by $\mathbb{\Sigma}^\text{H,D}_i = \frac{20}{3}\pi\eta a_i^3\mathbb{E}$.

For pairwise lubrication forces and torques acting between interacting particles $i$ and $j$ we start from the conventional representation given by~\citet{kim2013microhydrodynamics} as

\begin{equation}
 \begin{pmatrix} \bm{F}_{i,j}^\text{H,L} \\
 \bm{F}_{j,i}^\text{H,L} \\
 \bm{T}_{i,j}^\text{H,L} \\
 \bm{T}_{j,i}^\text{H,L}\\
 \mathbb{\Sigma}_{ii}^\text{H,L}\\
 \mathbb{\Sigma}_{j,i}^\text{H,L} \end{pmatrix}
  =\eta
  {\mathbb{R}}
   \begin{pmatrix}
   \bm{U}^\infty(\bm{x}_i)-\bm{U}_i \\
  \bm{U}^\infty(\bm{x}_j)-\bm{U}_j \\
 \bm{\Omega}^\infty-\bm{\Omega}_i \\
  \bm{\Omega}^\infty-\bm{\Omega}_j \\
   \mathbb{E}\\
   \mathbb{E}
   \end{pmatrix}\text{,}
\end{equation}
where ${\mathbb{R}}$ is the resistance matrix containing 
tensorial operations that linearly couple particle forces (torques) to
velocities (rotational velocities),
taking into account relative particle positions.
After some algebra and omitting terms that vanish with the size of the interparticle gap (see~\citet{Radhakrishnan} for details) one
can obtain the forces in a simplified pairwise form as
\begin{equation}
\begin{split}
	\bm{F}_{i,j}^\text{H,L}=-\bm{F}_{j,i}^\text{H,L}=& \left( X^A_{11}\mathbb{N}_{i,j}+Y^A_{11}\mathbb{T}_{i,j}\right)(\bm{U}_j-\bm{U}_i)\\
	&+Y^B_{11}(\bm{\Omega}_i\times\bm{n}_{i,j})\\
        &+Y^B_{21}(\bm{\Omega}_j\times\bm{n}_{i,j})\text{,}
\end{split}
\label{eqn:force_lub}
\end{equation}
where
$\bm{F}_{i,j}^\text{H,L}$ is the force acting on particle $i$ by particle $j$;
$\mathbb{N}=\bm{n}_{i,j}\otimes\bm{n}_{i,j}$ is a tensorial normal operator;
$\mathbb{T}=\mathbb{I} - \bm{n}_{i,j}\otimes\bm{n}_{i,j}$ is a tensorial projection operator;
$\bm{n}_{i,j}$ is the unit vector pointing from particle $i$ to particle $j$;
$\bm{U}_i$ is the velocity of particle $i$;
$\bm{\Omega}_i$ is the rotational velocity of particle $i$;
and $\mathbb{I}$ is the identity tensor in three dimensions.
The scalar prefactors $X$ and $Y$ encode the geometry of the interacting pair, 
namely the size of the interparticle gap and the size ratio of the interacting particles.
Their superscripts $A$, $B$ and subscripts $11$, $22$ are more appropriate to
the labelling convention used by~\citet{kim2013microhydrodynamics} but nonetheless
we retain them here for ease of referencing to that work.
The particle size ratio is written as $\beta=a_j/a_i$
and the dimensionless interparticle gap is $\xi = 2\left(|\bm{r}_{i,j}|-(a_i+a_j)\right)/(a_i+a_j)$.
The scalar prefactors are given by

\noindent\begin{minipage}{.5\linewidth}
\begin{equation}
X^A_{11} = 6\pi\eta a_i\left(\frac{2\beta^2}{(1+\beta)^3}\frac{1}{\xi} + \frac{\beta(1+7\beta+\beta^2)}{(5(1+\beta)^3)}\ln\left(\frac{1}{\xi}\right) \right)\text{,}
\end{equation}
\end{minipage}%
\begin{minipage}{.5\linewidth}
\begin{equation}
Y^A_{11} = 6\pi\eta a_i\left(\frac{4\beta(2+\beta+2\beta^2)}{15(1+\beta)^3}\ln\left(\frac{1}{\xi}\right) \right)\text{,}
\end{equation}
\end{minipage}
\noindent\begin{minipage}{.5\linewidth}
\begin{equation}
Y^B_{11} = -4\pi\eta a_i^2\left(\frac{\beta(4+\beta)}{5(1+\beta)^2}\ln\left(\frac{1}{\xi}\right)\right)\text{,}
\end{equation}
\end{minipage}%
\begin{minipage}{.5\linewidth}
\begin{equation}
Y^B_{21} = -4\pi\eta a_j^2\left(\frac{\beta^{-1}(4+\beta^{-1})}{5(1+\beta^{-1})^2}\ln\left(\frac{1}{\xi}\right)\right)\text{.}
\end{equation}
\end{minipage}\\

\noindent Meanwhile the torques on particle $i$ and $j$ as a result of their interaction with particles $j$ and $i$ respectively are written as
\begin{equation}
		\bm{T}_{i,j}^\text{H,L} = Y^B_{11}(\bm{U}_j-\bm{U}_i)\times\bm{n}_{i,j}-\mathbb{T}_{i,j}\left(Y^C_{11}\bm{\Omega}_i+Y^C_{12}\bm{\Omega}_j\right)\text{,}
  \label{eqn:torque_lub1}
\end{equation}
\begin{equation}
		\bm{T}_{j,i}^\text{H,L} = Y^B_{21}(\bm{U}_j-\bm{U}_i)\times\bm{n}_{i,j}-\mathbb{T}_{i,j}\left(Y^C_{21}\bm{\Omega}_i+Y^C_{22}\bm{\Omega}_j\right)\text{,}
  \label{eqn:torque_lub2}
\end{equation}
\noindent with scalar prefactors given by

\noindent\begin{minipage}{.5\linewidth}
\begin{equation}
Y^C_{11} = 8\pi \eta a_i^3\left(\frac{2\beta}{5(1+\beta)}\ln\left(\frac{1}{\xi}\right)\right)\text{,}\end{equation}
\end{minipage}%
\begin{minipage}{.5\linewidth}
\begin{equation}
Y^C_{12} = 8\pi \eta a_i^3\left(\frac{\beta^2}{10(1+\beta)}\ln\left(\frac{1}{\xi}\right)\right)\text{,}
\end{equation}
\end{minipage}
\noindent\begin{minipage}{.5\linewidth}
\begin{equation}
Y^C_{22} = 8\pi\eta a_j^3\left(\frac{2\beta^{-1}}{5(1+\beta^{-1})}\ln\left(\frac{1}{\xi}\right)\right)\text{,}
\end{equation}
\end{minipage}%
\begin{minipage}{.5\linewidth}
\begin{equation}
Y^C_{21} = 8\pi\eta a_j^3\left(\frac{\beta^{-2}}{10(1+\beta^{-1})}\ln\left(\frac{1}{\xi}\right)\right)\text{.}
\end{equation}
\end{minipage}\\

\noindent Similar expressions may be obtained for the elements of the hydrodynamic lubrication stress tensor,
though these can be shown to be equivalent (up to an order $\xi$ term in the normal stresses) to
the form used for the contact forces.
The contribution to the hydrodynamic stress coming from each pairwise interaction is thus given by $\mathbb{\Sigma}^{H,L}_{i,j} = -\bm{F}^\text{H,L}_{i,j} \otimes \bm{r}_{i,j}$.
To mitigate against divergence in the scalar prefactors at particle contacts (that is, where $\xi\to0$) we use $\xi_\text{eff}=10^{-3}$ in the calculation whenever $\xi< 10^{-3}$.
We do not calculate pairwise lubrication forces when particles are separated by gaps $\xi>0.05$,
having verified that this choice does not affect our conclusions.

\subsection{Brownian forces and torques}

To satisfy fluctuation-dissipation theorem,
we must produce Brownian forces that follow
\begin{equation}
\left\langle \mathcal{F}_B\otimes\mathcal{F}_B \right\rangle= \frac{2 k_b T}{\Delta t}{\mathcal{R}}\text{,} 
\end{equation}
where $\mathcal{F}_B$ is a list of the Brownian forces and torques, $\mathcal{R}$ is the overall resistance operator for the system (taking into account both one body and pairwise hydrodynamic dissipation that we describe separately below),
$k_bT$ is the thermal energy and $\Delta t$ is the computational timestep (discussed in more detail below).

For one-body Brownian forces we need 6 random numbers (i.e. two vectors in three-dimensional space  $\bm{\psi}_i$ and $\bm{\varphi}_i$) to satisfy the translational and rotational degrees of freedom of each particle $i$.
The elements of the random vectors $\bm{\psi}_i$, $\bm{\varphi}_i$ are drawn from a Gaussian distribution and satisfy  $\langle \varphi_\alpha\varphi_\beta \rangle = \langle \psi_\alpha \psi_\beta \rangle = \delta_{\alpha\beta}$ and they are uncorrelated with each other so that $\langle \varphi_\alpha \psi_\beta \rangle =0$.
The following forces and torques satisfy fluctuation-dissipation theorem (we label them as Brownian drag `B,D' to align with the hydrodynamic drag forces and torques defined above).
The one-body Brownian force and torque on particle $i$ are given by:
\begin{equation}
\bm{F}_i^{B,D} = \sqrt{\frac{2k_bT}{\Delta t}}\sqrt{6\pi \eta a_i}\bm{\psi}_i \text{,}
\label{eqn:force_brownian_drag}
\end{equation}
\begin{equation}
\bm{T}_i^{B,D} = \sqrt{\frac{2k_bT}{\Delta t}}\sqrt{8\pi \eta a_i^3}\bm{\varphi}_i\text{.}
\label{eqn:torque_brownian_drag}
\end{equation}
Averaging
$\langle\bm{F}_i^{B,D}\otimes\bm{F}_i^{B,D}\rangle$
and
$\langle\bm{T}_i^{B,D}\otimes\bm{T}_i^{B,D}\rangle$
over many realisations of the vectors $\bm{\psi}_i$ and $\bm{\varphi}_i$ leads,
respectively,
to
$\frac{2k_bT}{\Delta t}6\pi\eta a_i \mathbb{I}$
and
$\frac{2k_bT}{\Delta t}8\pi\eta a_i^3 \mathbb{I}$
as required (with $\mathbb{I}$ the identity matrix in three dimensions).

Pairwise Brownian forces and torques similarly require
two random vectors $\bm{\theta}_{i,j}$ and $\bm{\chi}_{i,j}$
(independent of $\bm{\psi}_i$ and $\bm{\varphi}_i$ but with the same properties)
to satisfy the relative translational and rotational motion
of two interacting particles.
The pairwise forces and torques
must be constructed in such a way that,
for particles $i$ and $j$,
averaging
$\langle \bm{F}_{i,j}^{B,L} \otimes \bm{F}_{i,j}^{B,L} \rangle$
and
$\langle \bm{T}_{i,j}^{B,L} \otimes \bm{T}_{i,j}^{B,L} \rangle$
over many realisations of 
$\bm{\theta}_{i,j}$ and $\bm{\chi}_{i,j}$
recovers the form of the pairwise hydrodynamic lubrication forces and torques described above.
Doing so,
which involves exploiting that the normal and projection operators present in the definition of the lubrication forces and torques are idempotent
(\emph{i.e.} $\langle\left(\mathbb{N}_{i,j}\bm{\theta}_{i,j}\right)\otimes \left(\mathbb{N}_{i,j}\bm{\theta}_{i,j}\right)\rangle = \mathbb{N}_{i,j}$)
and orthogonal
(\emph{i.e.} $\langle\left(\mathbb{N}_{i,j}\bm{\theta}_{i,j}\right)\otimes \left(\mathbb{T}_{i,j}\bm{\theta}_{i,j}\right)\rangle = 0$),
one obtains the following expressions for the pairwise Brownian force and torque
\begin{equation}
	\bm{F}_{i,j}^\text{B,L}=-\bm{F}_{j,i}^\text{B,L}=
 \sqrt{\frac{2k_bT}{\Delta t}}
 \left( \sqrt{X^A_{11}}
	\mathbb{N}_{i,j}+\sqrt{Y^A_{11}}\mathbb{T}_{i,j}\right)
 \bm{\theta}_{i,j}\text{,}
 \label{eqn:force_brownian_lub}
\end{equation}

\begin{equation}
		\bm{T}_{i,j}^\text{B,L}
  =
  \sqrt{\frac{2k_bT}{\Delta t}}\left(\frac{Y_B^{11}}{\sqrt{Y_A^{11}}}\bm{\theta}_{i,j}\times \bm{n}_{i,j}
		 +
		 \sqrt{Y_C^{11}-\frac{(Y_B^{11})^2}{Y_A^{11}}}\mathbb{T}\bm{\chi}_{i,j}
   \right)\text{,}
    \label{eqn:torque_brownian_lub1}
\end{equation}

\begin{equation}
		\bm{T}_{j,i}^\text{B,L}
  =
  \sqrt{\frac{2k_bT}{\Delta t}}\left(\frac{Y_B^{21}}{\sqrt{Y_A^{11}}}\bm{\theta}_{i,j}\times \bm{n}_{i,j}
		 +
		 \sqrt{Y_C^{22}-\frac{(Y_B^{21})^2}{Y_A^{11}}}\mathbb{T}\bm{\chi}_{i,j}
   \right)\text{.}
   \label{eqn:torque_brownian_lub2}
\end{equation}

\noindent Our model thus involves computing Equations~\ref{eqn:lang_force} and~\ref{eqn:lang_torque} to evaluate the trajectory of each particle,
subject to imposed forces given by
Equations~\ref{eqn:force_contact},~\ref{eqn:force_drag},~\ref{eqn:force_lub},~\ref{eqn:force_brownian_drag}, and~\ref{eqn:force_brownian_lub},
and torques given by Equations~\ref{eqn:torque_contact},~\ref{eqn:torque_drag},~\ref{eqn:torque_lub1},~\ref{eqn:torque_lub2},~\ref{eqn:torque_brownian_drag},~\ref{eqn:torque_brownian_lub1}
and~\ref{eqn:torque_brownian_lub2}.

\subsection{Brownian stress calculation}

One can similarly obtain from fluctuation-dissipation theorem an expression for the Brownian stress resulting from the pairwise interaction between particles $i$ and $j$ that averages over many realisations so that
$\langle\mathbb{\Sigma}^{B,L}_{i,j}\otimes\mathbb{\Sigma}^{B,L}_{i,j} \rangle$
recovers the form of the hydrodynamic lubrication stress,
but as described above this can similarly be shown to be equivalent to
$\mathbb{\Sigma}^{B,L}_{i,j} = -\bm{F}^\text{B,L}_{i,j} \otimes \bm{r}_{i,j}$.
Since the pairwise Brownian force term contains the normal operator $\mathbb{N}_{i,j}$
acting on the random vector $\bm{\theta}_{i,j}$,
one obtains a prefactor in the stress containing the dot product $\bm{n}_{i,j}\cdot\bm{\theta}_{i,j}$.
This quantity will always approach zero when averaged over many realisations of $\bm{\theta}_{i,j}$,
so that the Brownian stress computed in this way averages to zero.
Nonetheless,
particle pairs do experience non-zero Brownian forces acting at all timesteps that will
influence their trajectories so that the resulting contact and lubrication stresses will be altered by the presence of the Brownian forces.
Below we describe a method that allows us to estimate the contribution of Brownian motion to the overall stress.

It is important to note here that our method, in which particle inertia \emph{is} accounted for, is fundamentally different to other computational approaches, notably Stokesian Dynamics (SD)~\cite{ermak1978brownian,brady1988stokesian,bossis1989rheology} in which the trajectories are evolved with a timestep longer than the inertial one.
In the latter methods (see in particular~\citet{banchio2003accelerated}) the Brownian stress for the overall system is obtained as $\mathbb{\Sigma}^B = k_bT\nabla\cdot(\mathcal{R}_{SU}\cdot \mathcal{R}_{FU}^{-1})$,
in practice using a midpoint scheme in which the positions and velocities of
every particle are sampled at some increment of the overall timestep.
Here $\mathcal{R}_{SU}$ and $\mathcal{R}_{FU}$ represent parts of the overall
resistance matrix that couple, respectively, stresses to velocities and forces to velocities.
Our method described above is based on the Langevin equation
so that particle inertia is small but present,
and force balance is not strictly achieved at each timestep.
In order to obtain an estimate of the Brownian contribution to the stress,
we deploy a structural method that exploits the anisotropy of
the radial distribution function,
using the approach described by~\citet{brady1993rheological}.
The Brownian stress attributable to the pair $i,j$ can be written as
\begin{align}
        \mathbb{\Sigma}^B_{i,j} & = -n k_bT a \int_{S_2} (\bold{n}_{i,j}\otimes \bold{n}_{i,j}) p_{1/1}(\bold{x}_j| \bold{x}_i) \bold{dS}_2 \text{,}
\end{align}
where $p_{1/1}(\bold{x}_j| \bold{x}_i)$ is the probability density for finding a
particle at $\bold{x}_j$ given that there is a particle at $\bold{x}_i$, and $n=N/V$ is the number density of particles in the suspension
(where $V$ and $N$  are the system volume and particle number respectively).
The integral is over the surface of contact $S_2$ of two touching particles.

To compute this function we sum for each particle the diadic product of its unit vector with each of its neighbours within a thin shell $\Delta = 0.05a_i$, so that for a given configuration the Brownian contribution to the stress is~\cite{lin2016measuring}:
\begin{align}
        \mathbb{\Sigma}^{B} = -\frac{k_bT}{V}\sum_i \frac{a_i}{\Delta} \sum_{j \in \Delta}(\bold{n}_{i,j} \otimes \bold{n}_{i,j})\text{.}
\end{align}
The stress measured by this approach is not added to the hydrodynamic and contact stresses computed in our model,
but is available to provide insight into the role of Brownian motion in setting the overall material response.

%

\subsection{Additional simulation details}
We simulate $\mathcal{O}(10^3)$ spherical particles of radius $a$ and $1.4a$ (mixed approximately equally by volume) in a cubic periodic simulation box of length $L$. 
For each set of flow conditions we carried out between 10 and 800 realisations in order to obtain satisfactory ensemble averages.
The principle particle properties (these set the length, mass and time scales) are
the characteristic particle radius $a$ [length],
the particle density $\rho$ [mass/length$^3$]
(taken throughout to be equal to the fluid density so that the particles are neutrally buoyant),
and the particle normal stiffness $k_n$ [mass/time$^2$] (this has a tangential counterpart $k_t$).
With respect to these quantities, 1 time unit corresponds to the inverse frequency of a mass $\rho a^3=1$ on a linear spring with stiffness $k_n=1$.
The remaining material properties to be defined are
the fluid viscosity $\eta$ [mass/(length$\times$time)]
and
the particle-particle friction coefficient $\mu$ [dimensionless],
relevant for micron sized (and larger) particles.
The thermal energy scale in the system is set by $k_bT$.

The simulation box is deformed according to a specified $\nabla{\bm U}^\infty$.
For instance, when the only nonzero element of $\nabla{\bm U}^\infty$ is an off-diagonal (say $\dot{\gamma}$), shearing is applied by tilting the \emph{triclinic} box (at fixed volume) according to
$L_\text{xy}(t) = L_\text{xy}(t_0) + L\dot{\gamma}t$.
When the strain ($\gamma=\dot{\gamma}t$, with $t$ the time for which the simulation has run) reaches 0.5 in this example, the system is remapped to a strain of -0.5. This has no effect on the particle-particle forces or on the stress, and is simply a numerical tool to permit unbounded shear deformation while preventing the domain from becoming elongated in one axis~\cite{ness2021simulating}.
Reported in the following is the relative viscosity of the suspension $\eta_r=\Sigma_{xy}/\eta\dot{\gamma}$,
with $\Sigma_{xy}$ the shear component of the stress tensor,
$\dot{\gamma}$ the shear rate and $\eta$ the fluid viscosity.

\subsection{The timescales that appear in the simulation}

The full list of dimensional parameters taken as inputs to the model is then
$a$, $L$, $t$, $\rho$, $k_n$, $k_bT$, $\eta$ and  $\dot{\gamma}$.
Taking $a/L\ll1$ and $\dot{\gamma}t\gg1$,
dimensional analysis dictates that 
we require three non-dimensional groups to fully characterise this system.
In other words, a measured non-dimensional quantity e.g. the reduced viscosity $\eta_r = \Sigma_{xy}/\eta\dot{\gamma}$, can be a function of at most three non-dimensional control parameters.
This is in addition to non-dimensional inputs \emph{viz.} the volume fraction $\phi$ and the friction coefficient $\mu$.
Central to our work will be the study of viscosities as a function of Peclet number,
since this latter quantity will control the
colloidal to granular crossover.
It is desirous to choose the remaining two non-dimensional control parameters
such that particles are effectively \emph{hard} and \emph{non-inertial}.
To obtain an appropriate set of non-dimensional control parameters,
we consider the following list of timescales present in the model (in which we only include dimensional elements for simplicity):\\
\noindent\begin{minipage}{.25\linewidth}
\begin{equation}
\tau_\text{C}=\sqrt{\frac{\rho a^3}{k_n}}\text{,}
\label{contact_timescale}
\end{equation}
\end{minipage}%
\begin{minipage}{.25\linewidth}
\begin{equation}
\tau_\text{I}=\frac{\rho a^2}{\eta}\text{,}
\label{inertial_timescale}
\end{equation}
\end{minipage}
\begin{minipage}{.25\linewidth}
\begin{equation}
\tau_\text{B}=\frac{\eta a^3}{k_bT}\text{,}
\label{Brownain_timescale}
\end{equation}
\end{minipage}
\begin{minipage}{.25\linewidth}
\begin{equation}
\tau_\text{S}=\frac{1}{\dot\gamma}\text{.}
\end{equation}
\end{minipage}\\

The contact time $\tau_C$ is a characteristic time spent by two particles in contact (assuming contacts are describable as linear springs),
in the absence of other forces playing a role.
It is obtained by solving the following equation of motion for the overlap $\delta$ between contacting particles: $\rho a^3(d^2\delta/dt^2)=k_n\delta$.
The inertial relaxation time $\tau_I$ is the characteristic time taken for the velocity of a particle to reach that of the background fluid in the absence of other forces.
It is obtained by solving the following equation of motion for the velocity $v$ of a particle: $\rho a^3 (dv/dt) = \eta a v$.
The Brownian time $\tau_B$ is the characteristic time take for a particle to diffuse by a distance equal to its own radius under thermal motion in the absence of other forces.
The convective timescale $\tau_S$ is simply the inverse of the shear rate.
To resolve each of these timescales accurately within the simulation we chose the numerical timestep to be substantially smaller than the smallest of the timescales listed above. 
The Peclet number ($Pe$) described above is given by $6\pi\tau_B/\tau_S = 6\pi\eta a^3\dot{\gamma}/k_bT$,
and we vary this quantity across a broad range from 0.01 to 100000,
aiming to explore the colloidal to granular transition.

The contact time $\tau_C$ should be chosen to be sufficiently small that
overlaps between particles are orders of magnitude smaller than the particle radii, such that particles be considered hard spheres.
To do this we ensure throughout that $\tau_c$
is at least an order of magnitude smaller than the next smallest timescale.
The role of particle inertia can be expressed \emph{via} (i) a particle Reynolds number $\tau_I/\tau_S=\rho a^2\dot{\gamma}/\eta$,
and (ii) an inertia-diffusion ratio $\tau_I/\tau_B = \rho k_bT/\eta^2 a$.
Below we explore how small each of these quantities need to be set in order to
ensure inertia plays no significant role in the measured results.

\section{Results: interactions and diffusion}

\subsection{Two-particle simulations measuring the effective potential}
\begin{figure}
\includegraphics[trim = 0mm 0mm 15mm 0mm, clip,width=0.95\textwidth,page=2]{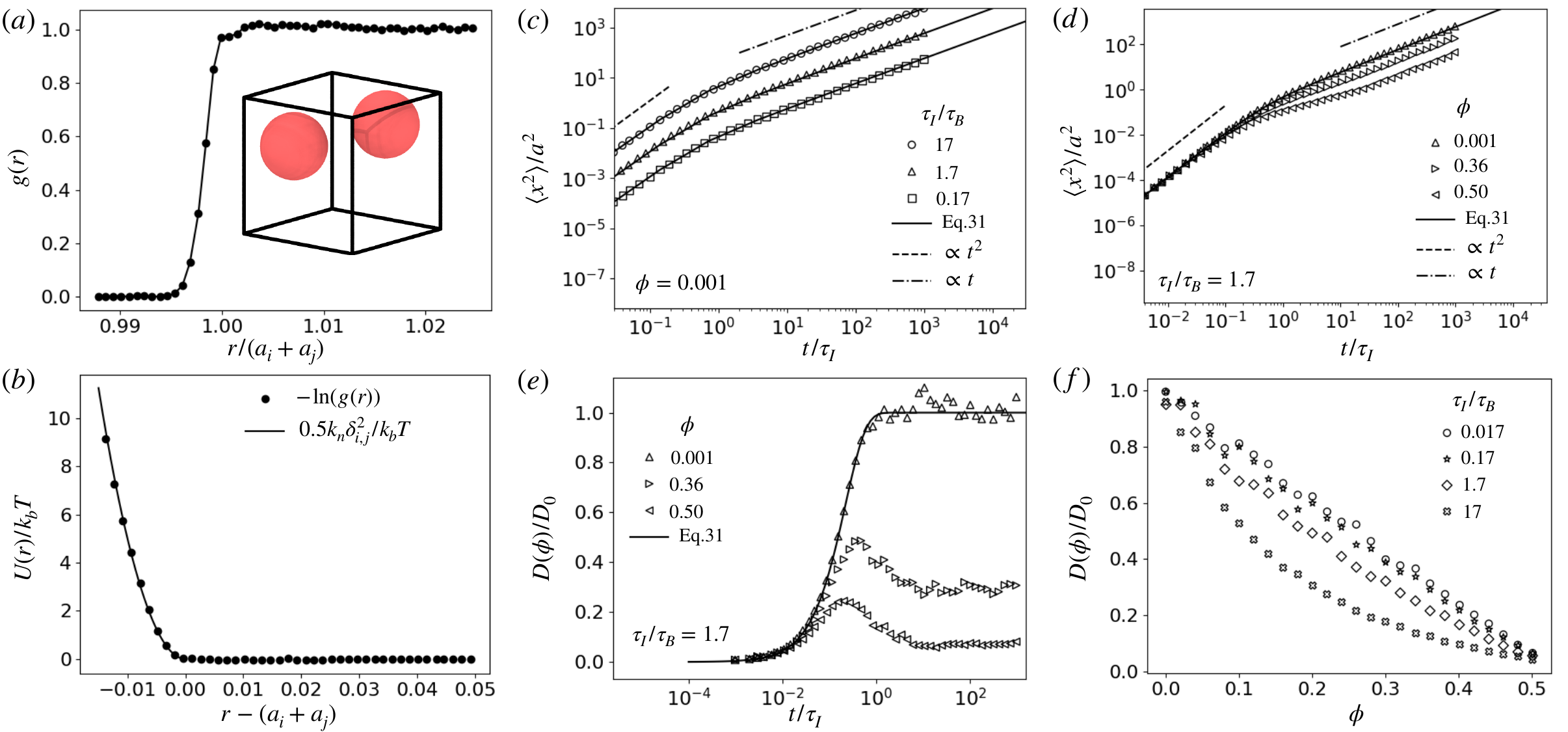}
    \caption{
    Evaluating the potential of mean force and the diffusion properties that arise from the particle-level forces described above, in the absence of shear flow.
    (a) The radial distribution function $g(r)$ (with $r = |\mathbf{r}_{i,j}|$) computed from a two particle simulation [Inset: snapshot of simulation];
    (b) Potential of mean force $U(r)$, showing measured result (points) and the input particle stiffness (solid line);
    (c)-(d) Mean squared displacement as a function of elapsed time for (c) three values of the timescale ratio $\tau_I/\tau_B$ at $\phi=0.001$; (d) three values of $\phi$ at $\tau_I/\tau_B=1.7$;
    (e) Diffusion coefficient as a function of elapsed time for a range of $\phi$ at $\tau_I/\tau_B=1.7$.
    The solid line in (c)-(e) represents the predictions of Equation~\ref{eqn:msd};
    (f) Long time diffusion coefficient at a broad range of $\phi$ and $\tau_I/\tau_B$.
    }
 \label{figure2}
\end{figure}

To evaluate the net pairwise potential resulting from the particle-level forces described above,
we carried out $\mathcal{O}(10^3)$ simulations of two particles
with radii $a$ in a cubic periodic box of length $4a$ (see snapshot in Figure~\ref{figure2}(a) Inset)
subject to all of the forces described above,
and with $\bm{U}^\infty=0$.
We calculate the radial distribution function $g(r)$ with $r=|\bm{r}_{i,j}|$ and averaged this across timesteps in the steady state and across all realisations (Figure~\ref{figure2}(a)),
then obtained the potential of mean force as $U(r)/k_bT = -\ln(g(r))$,
Figure~\ref{figure2}(b).
The result confirms that there is no net potential acting between particles when they are not in contact (\emph{i.e.} when $r>2a$),
so the lubrication and Brownian forces do not introduce an overall repulsion or attraction.
When particles are in contact ($r/(a_i+a_j)<1$) there is a steep repulsive potential
that,
as expected,
is related to the stiffness of our contacts defined above as $U(r)/k_bT = 0.5 k_n \delta_{i,j}^2$.
The model thus approximates a suspension of colloidal hard spheres,
in which the particle-particle interaction is zero and infinite for non-contacts and contacts respectively.

\subsection{Mean square displacement}
We next verify that our simulated particles follow statistically the anticipated trajectories
by computing their mean squared displacement (MSD) under various conditions.
An isolated particle with motion governed by the single body drag and Brownian forces described above is expected to follow a trajectory with a short-time ballistic part and a long-time diffusive part that leads to an overall MSD given by~\cite{lemons1997paul,hammond2017direct}:
\begin{equation}
    \langle x^2\rangle =2k_bT\frac{m}{\gamma^2}\left(\frac{\gamma}{m}t-1+e^{-\frac{\gamma}{m}t}\right)\text{,}
    \label{eqn:msd}
\end{equation}
with $m=(4/3)\pi\rho a^3$ and $\gamma=6\pi \eta a$.
This expression gives $\langle x^2\rangle\sim t^2$ and $\langle x^2\rangle \sim t$ at small and large times respectively.
It can equivalently be written in terms of our characteristic timescales defined above as:
\begin{equation}
    \langle x^2\rangle/a^2 =\frac{2}{27\pi   }\frac{\tau_I}{\tau_B}
    \left(
    4.5\frac{t}{\tau_I}
    -1
    +e^{-4.5\frac{t}{\tau_I}}\right)
    \text{,}
\end{equation}

Shown in Figure~\ref{figure2}(c) are MSDs for a dilute sample with $\phi=0.001$ in which pairwise particle-particle interactions are absent.
In terms of our model timescales, we set $\tau_S=\infty$ (\emph{i.e.} no shear);
$\tau_C=10^{-3}$; $\tau_I=10^{-1}$; and we vary $\tau_B$ to explore the behaviour at different temperatures.
We measure the elapsed time in units of $\tau_I$, so that the crossover from ballistic to diffusive behaviour begins in each case at $t/\tau_I\sim 1$.
As expected based on the expression above, increasing temperature (which decreases $\tau_B$) while keeping all other variables constant simply shifts the MSD result vertically with $\langle x^2\rangle \sim k_bT$.

We next calculate the MSD for a series of larger $\phi$,
with results shown in Figure~\ref{figure2}(d)-(e).
In all cases the particles follow a ballistic trajectory at short times
that is roughly independent of $\phi$.
The longer time behaviour shows a decreasing diffusion coefficient ($\mathcal{D}=d/dt(\langle x^2 \rangle$)) with increasing $\phi$,
a consequence of pairwise hydrodynamic and contact interactions resisting particle motion.
For all volume fractions below jamming $\mathcal{D}$ approaches a constant at long time scales,
confirming the presence of a diffusive regime. 

In order for inertia to play a negligible role in our model,
it is important for the diffusive timescale to be longer than the inertial relaxation one.
In other words,
the time taken for a particle velocity to relax to that of the background fluid should be much shorter than the time taken for the particle to diffuse by its own radius.
To understand quantitatively how to achieve this,
we measured $\mathcal{D}$ for varying $\tau_I/\tau_B$
across a broad range of $\phi$.
The normalized long time diffusion coefficient ($\mathcal{D}(\phi)/\mathcal{D}_0$) is shown in Figure~\ref{figure2}(f),
with
$\mathcal{D}_0=k_bT/\pi\eta a (=a^2/\tau_B)$.
Our result shows that when $\tau_I/\tau_B$ is smaller than $0.17$,
$D(\phi)/D_0$ becomes independent of temperature and follows a linearly decreasing trend.
This suggests a criteria for the maximum value of $\tau_I/\tau_B$,
which we check under shearing conditions in the following.

\section{Results: rheology}

In the following we first describe the need for substantial ensemble averaging, especially when Brownian motion dominates,
and we demonstrate the convergence of the measured rheology with the size of the sampling window.
We next go on to expose the role of particle inertia in our model under shear,
and establish the parameter range in which it can be assumed negligible.
We then present rheology data showing $\eta_r$ as
a function of $Pe$,
highlighting the breakdown of the individual contributions (hydrodynamic, contact and Brownian) and their variation with volume fraction.
We finally demonstrate the role of particle contact friction and a short-ranged repulsive potential.

\begin{figure}[b]
\includegraphics[trim = 0mm 80mm 0mm 0mm, clip,width=0.95\textwidth,page=3]{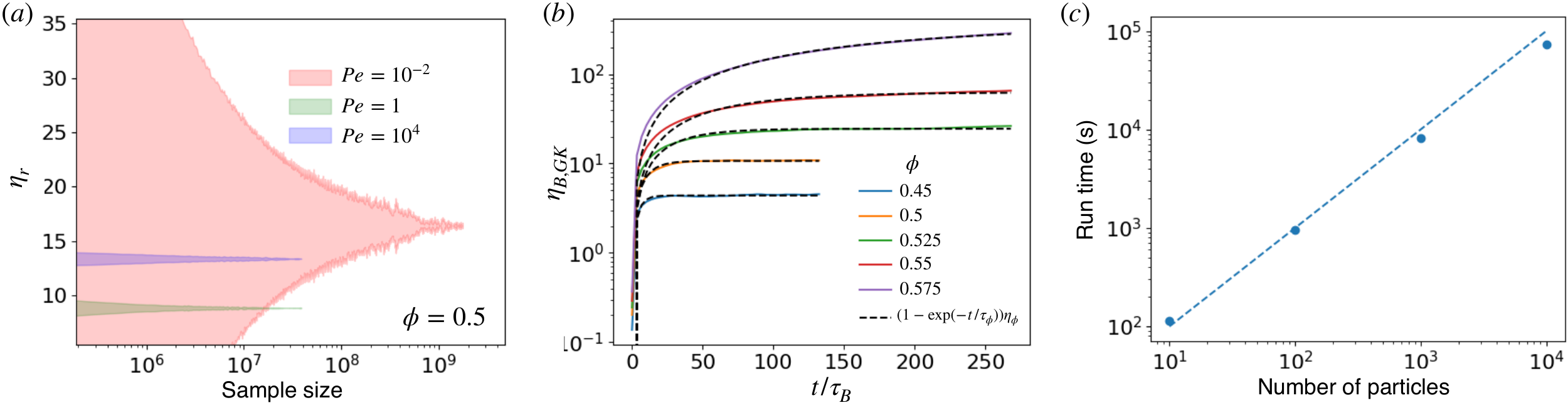}
    \caption{
Computing the suspension viscosity $\eta_r$ under sheared and non-sheared conditions, and the scaling of computational run time with system size.
Shown in (a) is the convergence of the measured $\eta_r$ as a function of the number of snapshots averaged over,
for $Pe=0.01$ (red),
$1$ (green)
and
$100$. (blue)
The noisy stress signal
when Brownian motion dominates necessitates large numbers of realisations.
In (b) is $\eta_{B,GK}$ measured \emph{via} the Green-Kubo relation taking the autocorrelation of the Brownian shear stress as input,
plotted as a function of the correlation time;
(c) Simulation run time versus number of particles for $\phi=0.5$, $Pe=1$, when running a serial compilation of LAMMPS on a single processor. We show data for a short simulation comprising $10^7$ timesteps.
    }
    \label{figure3}
\end{figure}

\subsection{Averaging method}

All of the rheology simulations described in the following were carried out with $\mathcal{O}(10^3)$ particles, comprising an
approximately equi-volume mixture of those with radius $a$ and $1.4a$. 
Given the comparatively small number of particles (compared to a real experimental system, for instance) and the random nature of the Brownian forces added to the system,
the stress signals output by a single simulation are extremely noisy, especially at low $Pe$.
(The same is true for inertia-free simulations~\cite{mari2015discontinuous}, though the error bars are rarely reported.)
Thus the number of realisations that must be averaged over to obtain smooth data and reliable estimates of the true rheology increases as $Pe$ is reduced.

Shown in Figure~\ref{figure3}(a) is the range of measured $\eta_r$ as a function of the number of steady state snapshots averaged over,
for 3 different $Pe$.
At low $Pe$ one must sample the system $\approx10^8$ times
to obtain a measurement of $\eta_r$ with standard deviation less than 10\%,
whereas for large $Pe$ $10^5$ samples are sufficient.
Importantly,
the time taken to reach steady state also differs
drastically with $Pe$.
For systems dominated by thermal fluctuation (i.e. low $Pe$)
the approach to steady state is set by the passage of Brownian time as opposed to the accumulated strain,
with systems at $\phi=0.54$ and below taking 3--4 Brownian times to reach steady state at $Pe=0.01$. For larger $\phi$ this timescale is stretched rapidly, likely due to the proximity of glassy physics.
At very large $Pe$,
meanwhile,
steady states are reached for strains $\dot{\gamma}t$ of 1--2~\cite{ness2016two}.

\subsection{Brownian stress at zero shear rate}
To obtain the Brownian contribution to the viscosity in the limit of zero shear rate,
we apply the Green-Kubo method~\cite{hansen2013theory} by calculating the time autocorrelation function of the shear stress,
taking as input the Brownian stress computed as described above, for unsheared simulations.
The Brownian viscosity is written as:
\begin{equation}
     \eta_{B,GK} = \frac{V}{k_bT} \int_0^\infty \langle {\Sigma}^B_{xy} (t+\Delta t){\Sigma}^B_{xy}(t)\rangle d\Delta t \text{,}
    \label{green kubo}
\end{equation}
where ${\Sigma}^B_{xy}$ is the shear component of the Brownian stress tensor.
The stress correlation decreases exponentially with increasing $\Delta t$ so that the Brownian viscosity can be modelled as 
$\eta_{B,GK}(t)  = \eta_{\phi} (1- e^{-\Delta t/\tau_{\phi}})$.
As shown in Figure~\ref{figure3}(b), the correlation time $\tau_{\phi}$ is short and weakly varying for $\phi<0.5$, so that $\eta_\phi$ can be measured using readily accessible data for which $\Delta t/\tau_\phi$ is large.
For $\phi>0.5$, however, $\tau_\phi$ grows quickly and we estimate $\eta_\phi$ by extrapolation. The rapid growth of the correlation time $\tau_\phi$ is likely indicative of a nearby glass transition, though we defer detailed analysis of this behaviour to future work.
By this approach we obtain an estimate of the Brownian contribution to the viscosity at zero shear rate,
which we discuss further in the following.

Given the large quantity of data required for obtaining smooth results,
it is worth considering the scaling of the simulation run time with the system size.
To estimate the scaling of the run time we carried out simulations with $N=10^1$, $10^2$, $10^3$, $10^4$ particles with $\phi=0.5$, running a serial build of LAMMPS on one core for $10^7$ timesteps. 
The result shown in Figure~\ref{figure3}(c) confirms that our simulation has complexity $\mathcal{O}$(N).

\subsection{The role of inertia}

\begin{figure}
\includegraphics[trim = 0mm 25mm 0mm 0mm, clip,width=0.8\textwidth,page=4]{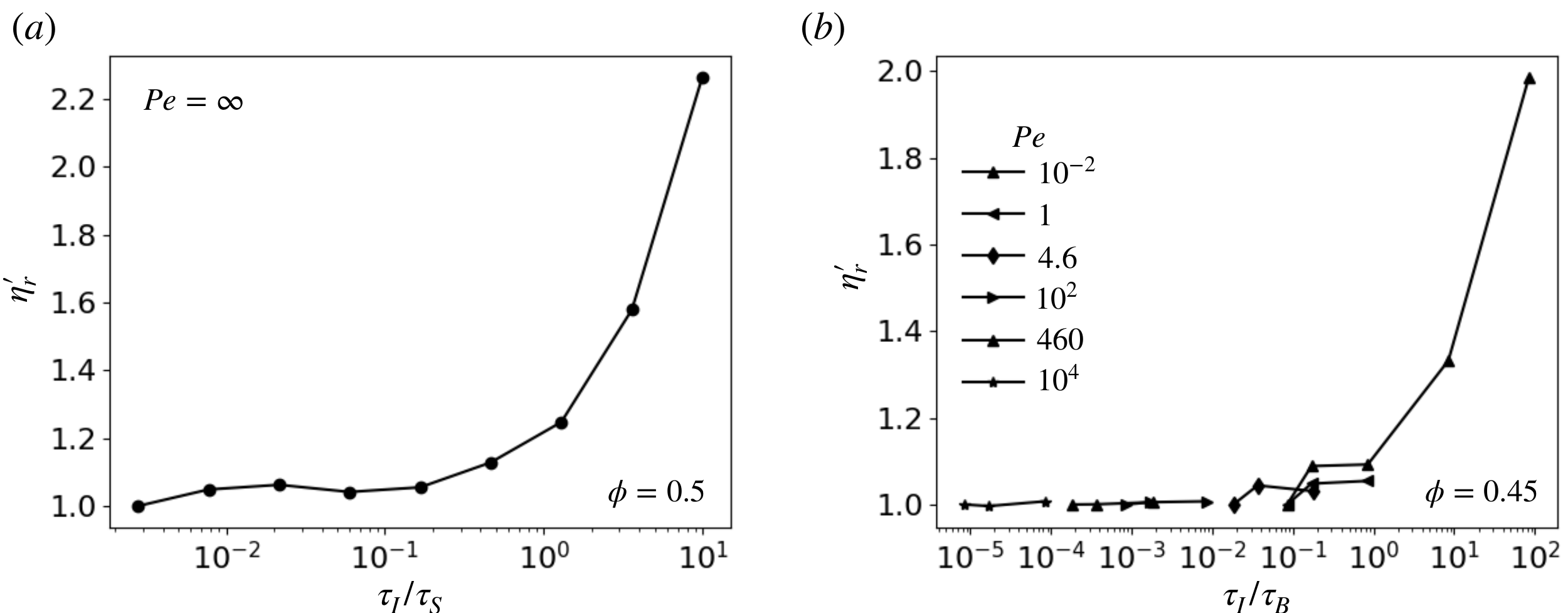}
    \caption{
    Establishing the parameter range in which particle inertia can be neglected.
    (a) Suspension viscosity $\eta_r$ (rescaled by the low $\tau_I/\tau_S$ value) plotted against the timescale ratio $\tau_I/\tau_S = \rho\dot{\gamma}a^2/\eta$ (the Stokes number) for $k_bT=0$ (so that $\tau_B=\infty$) and $\phi=0.55$, showing that our rheology results are rate-independent and therefore inertia-free for $\tau_I/\tau_S\lessapprox0.1$;
    (b) $\eta_r$ (rescaled by the low $\tau_I/\tau_B$ value) plotted against the timescale ratio $\tau_I/\tau_B=\rho k_bT/\eta^2a$ for $\tau_I/\tau_S<0.01$ and $\phi=0.45$. Shown are various values of $Pe$.
    }
    \label{figure4}
\end{figure}

\begin{figure}
\includegraphics[trim = 0mm 0mm 230mm 0mm, clip,width=0.7\textwidth,page=5]{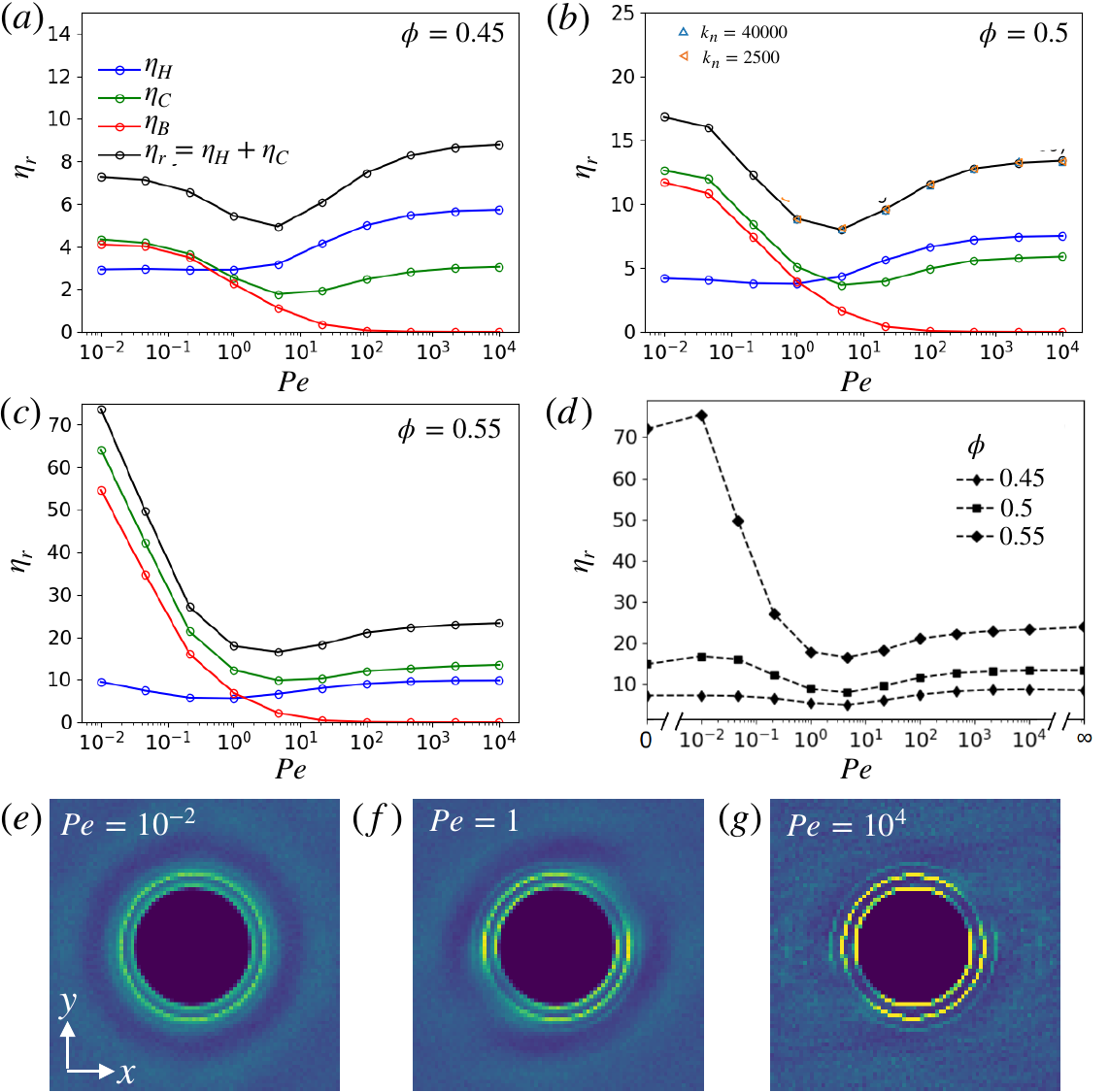}
    \caption{Rheology and microstructure of dense suspensions at the transition from Brownian to non-Brownian flow.
    Shown in (a)-(c) are the suspension viscosity $\eta_r$ as functions of $Pe$, for frictionless particles with
    (a) $\phi=0.45$;
    (b) $\phi=0.5$ (shown also are results for two additional values of $k_n$);
    (c) $\phi=0.55$,
    showing the contributions from contacts, hydrodynamics and Brownian forces.
    In (d) is the total viscosity as a function of $Pe$ and $\phi$.
    In (e)-(g) are slices through the three-dimensional radial distribution function $g(\bm{r}_{i,j})$ showing the flow-gradient ($xy$) plane under steady state simple shearing conditions for $\phi=0.5$ and
    (e) $Pe=0.01$;
    (f) $Pe=1$;
    (g) $Pe=10000$.
    }
    \label{figure5}
\end{figure}

The particle-particle contact timescale $\tau_C$ is
set sufficiently small that it does not compare to any other timescale in the system under any conditions,
so that particles can always be considered to be hard.
We verify this in Figure~\ref{figure5}(b) by showing that the relationship between $\eta_r$ and $Pe$ measured under different values of $\tau_C$ does not vary.
It is, however,
crucial that in varying $Pe$
one maintains acceptable values of $\tau_I/\tau_S$ and $\tau_I/\tau_B$.
To determine sufficiently small values of these two ratios so that inertia may be neglected,
we carried out two sets of simulations.
In the first we simulate shear flow with $\phi=0.55$ and $k_bT=0$ (so we don't need to consider the Brownian timescale $\tau_B$),
while varying the dimensionless shear rate $\tau_I/\tau_S$ from $5\times10^{-3}$ to $10$.
To be in the limit in which inertia is negligible,
we require a linear relation between the shear stress and the shear rate
\emph{i.e.} a Stokes flow.
In other words,
we are correctly simulating an inertia-free flow if $\eta_r$
is independent of $\tau_I/\tau_S$.
From Figure~\ref{figure4}(a) we can observe that
this holds for $\tau_I/\tau_S\lessapprox 10^{-1}$.
In what follows,
we therefore ensure that this inequality holds for all parameter sets.
Our result here is qualitatively consistent with prior simulations~\cite{trulsson2012transition}
and experiments~\cite{tapia2022viscous,madraki2020shear},
though the value of the Stokes number at the crossover
is apparently highly sensitive to system details.

In the second we simulate shear flow with $\tau_I/\tau_S<0.01$, $\phi=0.45$ and
at a range of $Pe$,
exploring the relative importance of inertia by varying the timescale ratio $\tau_I/\tau_B$.
This control parameter essentially sets the distance a particle will typically cover under ballistic motion.
In order for inertia to be negligible in the model,
we expect
that this distance should be at least an order of magnitude smaller than the particle size,
so that a typical Brownian \emph{kick} to a particle
does not lead it to collide with a distant neighbour.
From our result in Figure~\ref{figure2}(b)
we find that a ballistic to diffusive crossover
occurs at $\langle x^2 \rangle/a^2=0.01$ for $\tau_I/\tau_B=\mathcal{O}(10^{-2})$.
Our shear simulations (Figure~\ref{figure4}(b)) similarly
show that $\eta_r$ is a function of
$\tau_I/\tau_B$ only when the latter quantity is $>0.01$.
Therefore, in what follows we carry out simulations with $\tau_I/\tau_B<0.01$ and $\tau_I/\tau_S<0.1$.

\subsection{Flow curves}

Our main rheology results are presented in Figure~\ref{figure5}.
We simulated a broad range of $Pe$ ($10^{-2}-10^{4}$),
focussing on three different volume fractions $\phi$ (Figures~\ref{figure5}(a)-(c)) and
adhering to the constraints on $\tau_I$
obtained above.
To achieve this range of $Pe$ it was necessary to
vary both the shear rate $\dot{\gamma}$ and the thermal energy $k_bT$.
We present in Table~\ref{table1} a full list of the parameters used to generate the result in Figure~\ref{figure5}(a).
\begin{table}[b]
\begin{tabular}{ l|c|c|c|c|c|c } 
 $Pe$ & $\dot{\gamma}$ & $k_bT$ & $a$ & $\rho$ & $k_n$ & $\eta$ \\ \hline
 $\bm{0.01}$ &0.00009&0.017&1&0.1&10000&0.1 \\
 $\bm{0.046}$ &0.00042&0.017&1&0.1&10000&0.1 \\
 $\bm{0.22}$ &0.0019&0.017&1&0.1&10000&0.1 \\
 $\bm{1}$ &0.009&0.017&1&0.1&10000&0.1 \\
 $\bm{4.6}$ &0.009&0.0037&1&0.1&10000&0.1 \\
 $\bm{22}$ &0.009&0.00079&1&0.1&10000&0.1 \\
 $\bm{100}$ &0.009&0.00017&1&0.1&10000&0.1 \\
 $\bm{460}$ &0.009&0.000037&1&0.1&10000&0.1 \\
 $\bm{2200}$ &0.009&0.0000079&1&0.1&10000&0.1 \\
 $\bm{10000}$ &0.009&0.0000017&1&0.1&10000&0.1
\end{tabular}
\caption{
The parameters used to generate each data point in Figure~\ref{figure5}(a).
}
\label{table1}
\end{table}
In each rheology figure we break the overall viscosity down into its contributions from
hydrodynamic, contact and Brownian stresses.
The stresses obtained by taking the outer product of the pairwise vectors and forces
evaluated during the simulation run are the hydrodynamic one, the contact one, and the `instantaneous' Brownian stress.
The latter (not shown in Figure~\ref{figure5}),
as described earlier,
averages to zero so does not lead to a viscosity contribution.
The total stress (shown in black in Figure~\ref{figure5}(a)-(c)) is therefore just
the sum of the hydrodynamic and contact parts.

As a post-processing step we make an estimate of the effective Brownian
stress (approximating the one that would be measured in a Stokesian Dynamics simulation),
following the calculation based on structural anisotropy described earlier.
This gives us the red lines in Figure~\ref{figure5}(a)-(c).
Interestingly the Brownian stress maps quite closely to the contact stress for low $Pe$,
indicating that the surge in contact stress observed in this range
may be due to short-lived contacts induced by the Brownian kicks.
Indeed the formulation of the Brownian stress
is similar to that of the contact stress,
differing only in the presence of the contact overlap $\delta_{i,j}$ appearing in the latter.

Overall we find that the predicted rheology corresponds well with canonical results,
both in the experimental literature~\cite{de1985hard,laun1984rheological} and those obtained by Stokesian Dynamics simulation~\cite{foss2000structure} and similar numerical methods~\cite{mari2015discontinuous}.
At all volume fractions there is a shear thinning region for $Pe<1$
that gives way to shear thickening beyond $Pe>1$,
with the $\eta_r$ values at large $Pe$ tending towards those reported for non-Brownian
suspensions under a very similar numerical framework~\cite{cheal2018rheology}.
For $\phi=0.45$ and $\phi=0.5$ we observe a low $Pe$ plateau,
whereas at $\phi=0.55$, $\eta_r$ apparently continues to increase with decreasing $Pe$.
The latter effect is perhaps an artefact of proximity to a glass transition,
though we defer a more detailed study of this effect to future work due to the diverging timescales involved.
The hydrodynamic stress increases weakly with increasing $Pe$,
whereas the contact stress qualitatively follows the overall stress in its shape.
The increase in contact viscosity at high $Pe$ may be attributed to the onset of contact force chains
as the system approaches the non-Browian limit and can be considered granular~\cite{lin2015hydrodynamic},
while at low $Pe$ it is related to the Brownian forces as described above.

Shown in Figure~\ref{figure5}(d)-(f) are slices through the three dimensional radial distribution function $g(\bm{r}_{i,j})$,
showing the flow-gradient ($xy$) plane at $Pe=0.01$, $Pe=1$ and $Pe=10^4$.
The general shape of the pairwise distributions is consistent with literature data~\cite{foss2000brownian},
showing increased anisotropy with increasing $Pe$ and a sharpening of the peaks 
at $a + a$, $a+1.4a$ and $1.4a + 1.4a$.

\begin{figure}[b]
\includegraphics[trim = 0mm 82mm 0mm 0mm, clip,width=0.95\textwidth,page=6]{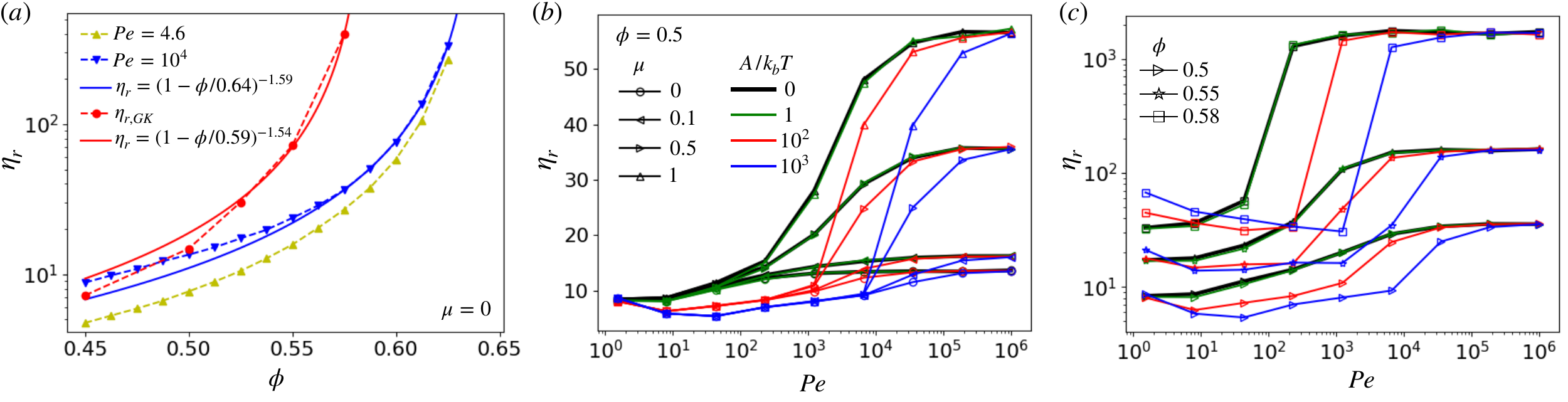}
    \caption{
    The viscosity variation with volume fraction and particle-particle contact friction.
    (a)
    Variation of $\eta_r$ with volume fraction $\phi$ at three $Pe$, showing fits to $\eta_r=(1-\phi/\phi_J)^{-\lambda}$;
    (b)
    $\eta_r$ as a function of $Pe$ for several particle-particle friction coefficients $\mu$
    and repulsive force magnitudes $A/k_bT$
    at a volume fraction of $\phi=0.5$;
    (c)
    $\eta_r$ as a function of $Pe$ for several $\phi$ and $A/k_bT$, with $\mu = 0.5$.
    The colour legend in (b) refers also to (c).
    }
    \label{figure6}
\end{figure}

\subsection{Viscosity variation with volume fraction}
In order to understand better the limiting behaviour at small $Pe$,
we determine the behaviour of $\eta_r$ as a function of $\phi$.
To do so we first evaluate the Brownian contribution to the zero shear viscosity using the Green-Kubo method described above.
To obtain an estimate of the full viscosity,
we take the value of the hydrodynamic viscosity at the smallest (non-zero) measured $Pe$,
and add this to the Green-Kubo prediction of the Brownian stress (assuming the latter to be a good proxy for the contact stress, as was assumed by~\citet{brady1993rheological} and is supported by our simulation data in Figure~\ref{figure5}).
Doing so at a range of $\phi$,
and comparing the result to the minimum $\eta_r$ measured at $Pe=4.6$ for each $\phi$ as well as the large $Pe$ limit,
we obtain Figure~\ref{figure6}(a).

In both the low and high $Pe$ limits, we find that $\eta_r$,
particularly at large $\phi$,
can be fit relatively well with a simple relation as $\eta_r\approx(1-\phi/\phi_J)^{-\lambda}$,
with $\phi_J(Pe\to0)=0.587$
and
$\phi_J(Pe=10^4)=0.642$
(and $\lambda\approx1.5$, similar to~\citet{mari2014shear}).
At intermediate $Pe$, $\eta_r$ is reduced relative to its value in the the non-Brownian limit, and the value of $\phi_J$ is marginally increased.
The large $Pe$ value of $\phi_J$ will be highly sensitive to details of the particle-particle contact interaction,
especially the presence of a static friction coefficient as we have reported elsewhere~\cite{cheal2018rheology,singh2020shear}.
In particular,
for large friction coefficients the large $Pe$ value of $\phi_J$ (usually denoted $\phi_m$) will likely drop below the low $Pe$ value.
In this scenario one expects flow curves near jamming to be diverging at both low and high $Pe$,
with finite $\eta_r$ at intermediate $Pe$.
We leave this complexity to be explored in future work,
and in the following we examine the role of friction for a small range of $\phi$.

\subsection{Role of particle-particle friction and short-ranged repulsion}

In the context of experimental work by~\citet{guy2015towards},
it is important to consider the role of particle friction at the colloidal-to-granular transition.
Since granular particles are large, micron size objects they will likely have
a static friction coefficient,
which may constitute both sliding and rolling components~\cite{singh2020shear,blair2022shear}.
So far we have only considered a model system of frictionless particles.
It is well-established that the presence of static sliding friction means that
each particle-particle contact will constrain more than one degree of freedom of each particle,
so that for large friction coefficients (in practice $\mu\gtrapprox0.5$) a rigid packing can be obtained with a per particle contact number of $\approx4$ (as opposed to 6 for frictionless spheres),
with limiting volume fraction~$\phi_m\approx0.57$.
In Figure~\ref{figure6}(b) we report rheology predictions from simulations of suspensions with 
a range of particle-particle friction coefficients $\mu$ (black data),
demonstrating that the presence of friction leads to a dramatic increase in $\eta_r$ at large $Pe$.
This behaviour, and its sensitivity to $\phi$ demonstrated in Figure~\ref{figure6}(c) (black data), is qualitatively consistent with the large literature on friction-driven shear thickening e.g.~\citet{mari2014shear}.
Notably, $\eta_r$ at lower $Pe$ is unaffected by friction,
suggesting that Brownian forces suppress the
mobilisation of static friction for all $\mu$, at least at $\phi=0.5$.
In this respect the Brownian forces act analogously to a weak repulsive potential, inhibiting the formation of sustained particle contacts
and rendering the suspension effectively frictionless even when $\mu>0$.
This leads to a bulk viscosity with rate dependence qualitatively similar to that of shear thickening suspensions
with load-activated friction describable by the canonical model of~\citet{wyart2014discontinuous}.

Importantly, though, is it not clear that the shear thickening transition, when controlled by Brownian motion, is governed by a single stress scale.
In particular, the range of $Pe$ over which the transition happens in Figure~\ref{figure6}(b) (black data) is rather broad (occurring over 4-5 orders of magnitude in $Pe$), especially when compared to~\citet{mari2014shear} in which the transition takes at most 2 orders of magnitude in shear rate.
To explore this we introduce a short ranged repulsive force defined by
\begin{equation}
    \bm{F}_{i,j}^\text{R}=\frac{A}{\kappa}\exp\left(\frac{(a_i+a_j)-|\bm{r}_{i,j}|}{\kappa}\right) \bm{n}_{i,j} \text{,}
\end{equation}
with $\kappa=0.01(a_i+a_j)$.
We show results of this model for $A/k_bT=0, 1, 10^2, 10^3$ in Figure~\ref{figure6}(b) and for several $\phi$ at $\mu=0.5$ in Figure~\ref{figure6}(c).
Introducing a sufficiently large repulsive force scale (in practice we required $A/k_bT\approx100$) narrows the range of $Pe$ over which shear thickening occurs, and shifts the transition to larger $Pe$.
This result suggests not only an additive effect of Brownian and repulsive forces as reported by~\citet{mari2015discontinuous},
but rather a qualitative change in the functionality of $\eta_r$ with $Pe$ when the onset of contacts is set by the magnitude of Brownian or repulsive forces.
Examining the subtly in more detail is a promising area in which our model might be deployed.
Thus with the introduction of particle-particle friction and a short ranged repulsive force we
can control in our model the position and extent of shear thickening,
providing a flexible starting point from which to make predictions of the rheology in more specific contexts.

\section{Concluding remarks}

In conclusion, we have implemented a minimal numerical model for the rheology of dense suspensions that incorporates sufficient microscopic physics to predict the colloidal to granular crossover as a function of $Pe$.
The model is implemented in LAMMPS~\cite{plimpton1995fast} so that its run time scales linearly with the number of particles.
The Brownian component of our model differs from that in Stokesian Dynamics in that we resolve the fluctations at a much shorter, inertial timescale. The naively-calculated Brownian stress therefore averages to zero over realisations and instead we compute an estimation of the Brownian contribution to the stress based on the structural statistics measured from the simulation. This stress follows closely the contact stress that we measure directly from the pairwise forces and relative positions.
The model predicts shear thinning at low $Pe$, with a low $Pe$ plateau (in some cases) that increases with volume fraction.
At larger $Pe$ a Brownian regime gives way to a contact dominated regime in which particle-particle interactions proliferate and friction (if present) becomes important.
In this latter regime shear thickening is observed even for zero particle friction,
though its extent increases with increasing friction coefficient.
We finally introduced into our model a short range repulsive force,
a crucial prerequisite for shear thickening
in the paradigmatic model of non-Brownian suspensions~\cite{mari2014shear}.
This
keeps particles separated and inhibits the contact contribution to the stress,
thus broadening the intermediate $Pe$ viscosity plateau (as observed by~\citet{cwalina2016rheology}) or equivalently shifting the value of $Pe$
at which particle contacts become important.

We have focussed in this article on steady, simple shear rheology.
Broadening the work to inhomogeneous conditions (such as those described by~\citet{gillissen2020modeling})
and to dynamic simple shear to measure the frequency-dependent (and indeed amplitude dependent~\cite{ness2017oscillatory}) response
are promising lines of future research that will provide additional scope for constitutive model development and for validation against experimental data.

In future we anticipate deploying our code in mixed systems, in which one population of particles are Brownian and another are non-Brownian~\cite{cwalina2016rheology}.
This is motivated by numerous real world examples such as geophysical flows and many scenarios in chemical engineering and manufacturing.
In such systems the small, Brownian particles (in some industries these are referred to as superplasticizers) will simultaneously contribute a Brownian stress but improve the efficiency of packing,
so that their overall effect on the rheology is non-trivial and likely to be non-monotonic and $Pe$-dependent.
Mapping out this complexity as functions of the small and large particle sizes and their relative numbers requires a tractable numerical model,
and will likely rely on the implementation of more advanced neighbour listing algorithms such as those by~\citet{shire2021simulations}.
Extending this further to systems with continuous, broad size distributions remains on open challenge~\cite{mwasame2016modeling}.

\section{Acknowledgements}
Codes and scripts necessary to reproduce the results reported in this article are available on request.
C.N. acknowledges support from the
Royal Academy of Engineering under the Research Fellowship scheme
and from the Leverhulme Trust under Research Project Grant RPG-2022-095.
We thank John Brady, Jeff Morris, Aleksander Donev, Emanuela Del Gado, Abhay Goyal, Anthony Ge, Romain Mari and Ryohei Seto for useful discussions.

\bibliography{library}
\end{document}